\documentclass{article}
\usepackage{graphics}
\usepackage{setspace}

\begin{document}
%\begin{frontmatter}
\title{Raman Tensor Calculation for Magnesium Phthalocyanine}
\author{
 Jaroslav T\'{o}bik$^{a}$, Erio Tosatti$^{a,b}$ \\
\small $^a$International School for Advanced Studies (SISSA),\\
\small and INFM Democritos National Simulation Center, Via Beirut 2, I-34014 Trieste, Italy\\
\small $^b$The Abdus Salam International Centre for Theoretical Physics (ICTP),\\
\small strada Costiera 11, 34100 Trieste, Italy  }
\date{}
\maketitle
\begin{abstract}
We present ab-initio density functional (DFT) calculations of the 
vibrational spectra of neutral Magnesium phthalocyanine (MgPc) molecule
and of its Raman scattering intensities.
\end{abstract}
%\end{frontmatter}
\section{Introduction}
Organic materials are attractive for electronic industry, because they promise 
large versatility based on organic chemistry, low price based on very common 
elements used as building blocks and interesting mechanical properties. 
There is quite a vigorous 
research activity in organic materials for electrical applications - from
insulators trough semiconductors to conductors and even superconductors.
Here we are concerned with metal phthalocyanines (MPc).
Electronic structure calculations \cite{Liao} and recent experimental 
observation of the possibility to dope
these materials by adding electrons \cite{Morpurgo} resembles somewhat the 
situation 
of fullerenes. In our previous paper we speculated about possible 
phase diagrams for electron doped MgPc \cite{Erio-PRL}. While these
possibilities are under active experimental considerations, we noticed
that for the commonest experimental diagnostic, namely Raman scattering
there is no reference calculation of either modes nor Raman intensities
(with exception of ZnPc \cite{Tackley-PCCP1}).
In this paper we focused our interest on the vibrational properties of MgPc, 
which we wish to study ab-initio. While our final aim will be to compute
spectra of doped and undoped molecules, this work will be restricted
to neutral undoped MgPc, which has not been studied so far.

\section{Technical details.}
All our calculations were done using the PWscf software package \cite{PWSCF}, 
which is a plane-wave basis set DFT implementation. We used 
the LDA approximation 
with Slater approximation for exchange and Perdew-Zunger functional 
for correlations effects. 
In order to lower the plane wave energy cut-off we used the non-local 
RRKJ3 ultra--soft pseudopotentials \cite{RRKJ3}.
The kinetic energy cut-off was 35Ry for the wave functions basis set and 280Ry 
for the charge density basis set. The plane-wave basis set assumed periodic 
boundary conditions.  The unit cell had dimensions 
21$\times$21$\times$11{\AA}, therefore including enough vacuum 
to represent the isolated molecule.

By energy optimization we found the equilibrium molecular structure 
with structural 
parameters as
in Table \ref{tab-struct}. The molecule has $D_{4h}$ symmetry group 
and is planar. 
The electronic structure 
agrees well with former calculations of Liao et. al.\cite{Liao}. 
Some levels near the HOMO--LUMO gap are listed 
in the table \ref{tab-electrons}.

We calculated the vibrational spectra of neutral, undoped MgPc by means of 
the density functional perturbation theory \cite{Baroni-RMP}.
The dynamical matrix was calculated in Cartesian coordinates.
In the isolated molecule with $N$ atoms there are $3N-6$ genuine
vibrations while 6 modes corresponds to translation--rotational 
degrees of freedom with zero frequency. 
In reality due to periodic images of molecules there is a weak 
virtual interaction which can cause "libration" of the molecule at nonzero, 
even imaginary frequency.
We eliminated these modes by transforming the dynamical matrix 
to internal coordinates and setting the corresponding dynamical 
matrix elements to zero.
We checked that after this procedure we obtained eigenvectors 
of correct symmetry.
Assignment of irreducible representation to all eigenvectors was done 
by projection on symmetry adapted bases of all linearly 
independent atomic displacement.
For $N=57$ atoms there are $165$ vibrations 
$$
\Gamma_{vib}=14A_{1g}+13A_{2g}+14B_{1g}+14B_{2g}+13E_g+6A_{1u}+8A_{2u}+7B_{1u}+
7B_2u+28E_u
$$
where $E_g$ and $E_u$ are two--fold degenerated modes. 
The calculated frequencies are summarized in Fig. \ref{fig-spectrum}.

The Raman intensity is related to the change of molecular polarizability 
due to the deformation introduced by the vibration. 
We calculated the polarizability derivatives with respect to deformations 
in the static limit. 

With each vibrational mode $q_i$ is associated a Raman tensor $T^i$
given by 
\begin{equation}
T^i_{\mu \nu}=\frac{d\alpha_{\mu \nu}}{dq_i}
\label{raman-ten}
\end{equation}
 Here $\mu$, $\nu$ are Cartesian indices, $\alpha_{\mu \nu}$ 
is the polarizability tensor defined as
$\alpha_{\mu\nu}=\frac{\partial ^2U}{\partial E_{\mu} \partial E_{\nu}}$, 
$U$ is the total energy and $\vec{E}$ an external electric field. Therefore one has to calculate
third derivatives of total energy 
$\frac{\partial^3 U}{\partial E_{\mu} \partial E_{\nu} \partial x_l}$. 
The energy derivatives with respect to atomic displacements are forces 
acting on atomic nuclei, and are calculated via the Hellman-Feynman theorem.
We extracted the force dependence on electric field 
by numerical differences. Applied fields were $\pm 1\times 10^{-3}$, 
$\pm 2\times 10^{-3}$, $\pm 4\times 10^{-3}$ a.u.. Details about application of
electric field within periodic boundary conditions are the same as in 
Ref. \cite{Andrea-JCP}. There are alternative methods 
for the Raman tensor calculation using linear response theory 
\cite{Deinzer,Lazzeri}. However numerical derivative method with finite 
field is computationaly less demanding for the present case, where moreover 
the implementation is very simple.

From the Raman tensor Eq.(\ref{raman-ten}) 
we calculated for each mode the gas phase (angle averaged)
Raman scattering cross section 
following \cite{Pederson,Long-book} in the form:
\begin{eqnarray}
\frac{d \sigma _i}{d \Omega}&=& \frac{(2\pi\nu_0)^4}{c^4}\frac{h(n_i+1)}{8\pi\nu_i}\frac{45{\alpha'}_i^2+7{\gamma'}_i^2}{45} \label{ram-cross}\\ 
\alpha'_i&=&\frac{1}{3}\left(T^i_{xx}+T^i_{yy}+T^i_{zz}\right) \nonumber\\
{\gamma'}_i^2&=&\frac{1}{2}\lbrace 
\left( T^i_{xx}-T^i_{yy} \right)^2 + \left( T^i_{xx}-T^i_{zz} \right)^2 +
\left( T^i_{yy}-T^i_{zz} \right)^2 + 
6\lbrack \left(T^i_{xy}\right)^2 + \left(T^i_{xz}\right)^2 +\left(T^i_{yz}\right)^2 \rbrack
\rbrace \nonumber \\
n_i&=&{\left[ \exp \left(\frac{h\nu_i}{kT}\right) -1\right]}^{-1}
\nonumber 
\end{eqnarray}
$\nu_0$ is frequency of incident light, $n_i$ is equilibrium occupation 
number for the initial vibrational state $i$ at the given temperature $T$.
The $\alpha '$ and ${\gamma '}^2$ are 
isotropic and anisotropic parts of the Raman tensor. 
Both are invariant under rotations. Formula (\ref{ram-cross}) is valid 
for the most common experimental setup when incident ray, observed ray, 
and incident light polarization of electric field are perpendicular 
to each other. 

\begin{figure}[h]
{\par\centering \resizebox*{0.7\textwidth}{!}{\includegraphics{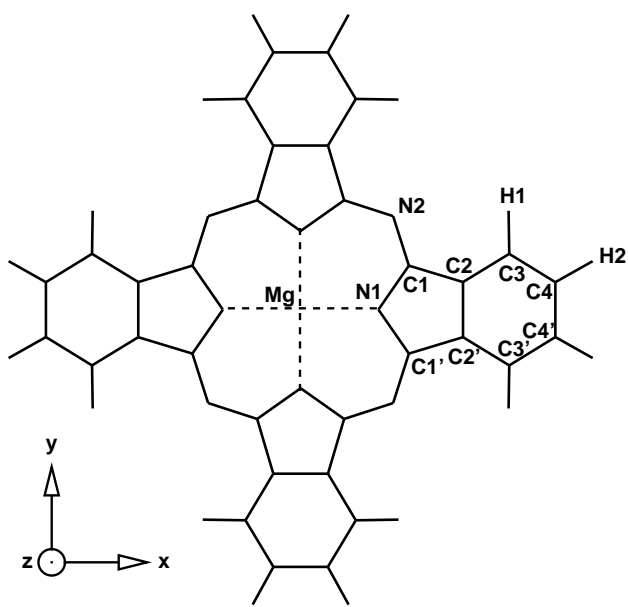}}\par}
\caption{Schematic structure of the MgPc molecule. Orientation of coordinate system is important for definition of the Raman tensors by Eq. (\ref{ram-tensors}).}
\label{fig-mol-scheme}
\end{figure}

\begin{figure}
{\par\centering \resizebox*{!}{0.9\textheight}{\includegraphics{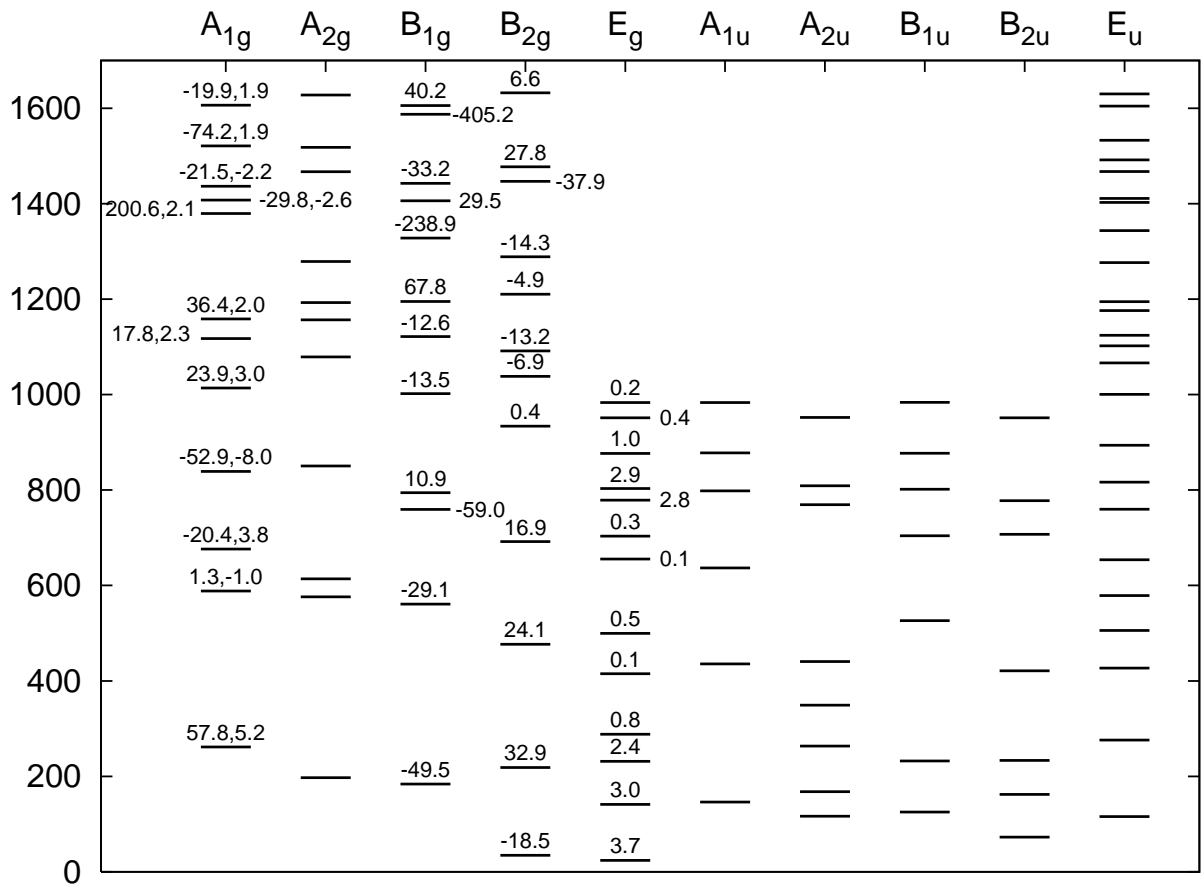}}\par}
\caption{Frequencies and independent Raman tensor components of neutral MgPc. 
Frequencies are in cm$^{-1}$, Raman tensor components in atomic units (distance in Bohr radii $a_0$, energy in Rydbergs, and electric fields units $e/a_0$). 
Note that $A_{1g}$ has two independent 
components, the first is $a$, and the second $b$ 
following the notation in Eq.(\ref{ram-tensors}).}
\label{fig-spectrum}
\end{figure}

The Raman tensor of a particular mode  
must belong to the same irreducible representation as that vibration mode. 
In a coordinate system connected to the molecule as indicated 
on Fig. (\ref{fig-mol-scheme}) all symmetric tensors of second rank 
are divided according to irreducible representations in the form:
\begin{equation}
A_{1g}:\left(\begin{array}{ccc}
a & 0 & 0 \\
0 & a & 0 \\
0 & 0 & b 
\end{array} \right), \quad 
B_{1g}:\left(\begin{array}{ccc}
c & 0 & 0 \\
0 & -c & 0 \\
0 & 0 & 0
\end{array} \right), \quad
B_{2g}:\left(\begin{array}{ccc}
0 & d & 0 \\
d & 0 & 0 \\
0 & 0 & 0
\end{array} \right), \quad 
E_g:\left(\begin{array}{ccc}
0 & 0 & e \\
0 & 0 & f \\
e & f & 0 
\end{array} \right)
\label{ram-tensors}
\end{equation} 
Traceless irreducible representations possess only the anisotropic part, 
while the $A_{1g}$ mode has both parts.
Weights in front of $\alpha'$  and $\gamma'$  
in formula \ref{ram-cross} (45 and 7 respectively) can
be changed by observation geometry, thus providing a way to (partially) 
assign symmetry to vibration eigenmodes 
(see for example Ref. \cite{Long-book}). 
We note that for the $E_g$ representation it is possible to define 
a single independent parameter $\sqrt{e^2+f^2}$ because of the degeneracy
and orthogonality of degenerated modes.

\section{Results and discussion}
In Table \ref{tab-struct} we give our LDA optimized structure of MgPc. 
In comparison with data in references \cite{Liao,Ruan} we see a 
reasonable agreement.
The difference can be attributed to different density functional 
(GGA in Ref. \cite{Liao}). LDA is known to slightly overbind in covalent bonds. 
Kohn-Sham orbitals energies (Tab. \ref{tab-electrons}) are given relative to
vacuum zero taken as the SCF electrostatic potential far from the molecule. 
They are also in good agreement with Ref. \cite{Liao} as far as we can read off
values from fig. 2 of Ref. \cite{Liao}. 

Vibrational spectra of various metal-phthalocyanines were experimentally 
measured \cite{Tackley-PCCP2} by means of Raman spectroscopy. 
In the case of ZnPc there is also a DFT calculation of the vibrational 
and Raman spectra \cite{Tackley-PCCP1}. 
In general the most intense peak was found to be in the interval 
1500-1550cm$^{-1}$ depending on the central atom. 

\begin{figure}
{\par\centering\resizebox*{1.0\textwidth}{!}{\rotatebox{270}{\includegraphics{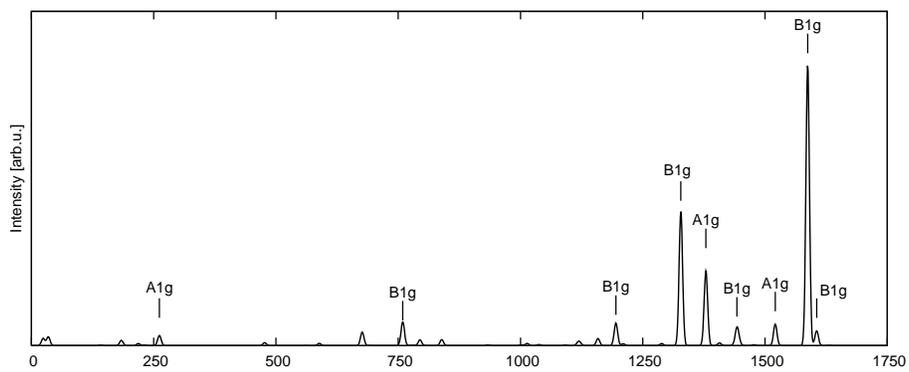}}}\par}
\caption{Calculated Raman spectrum of MgPc for very low excitation frequency and low temperature.}
\label{fig-raman}
\end{figure}

Our calculated vibration spectrum of MgPc is on Fig.(\ref{fig-spectrum}). 
Above the frequency line for Raman active modes we write the  corresponding 
non-zero component of Raman tensor as defined by Eq.(\ref{ram-tensors}). 
The simulated Raman spectrum according to formula \ref{ram-cross} 
is shown on Fig.(\ref{fig-raman}). 
To simulate spectrum for each intensity and frequency we added
Gaussian with the spread $\sigma$=5cm$^{-1}$.
We considered the low temperature limit ($n_i=0$) and low incident light 
frequency limit. 
Our most intense Raman scattering mode is at 1587cm$^-1$ with 
symmetry $B_{1g}$. This is similar to ZnPc
which is at 1516cm$^{-1}$ with symmetry $B_1$. 

The results of the presented calculation should be compared with Raman data
of MgPc, as soon as available. It will also serve as a starting point for future work on the electron doped molecules \cite{Morpurgo}.

\section*{Acknowledgments}
We are grateful to A.~Dal~Corso for his help and assistance and to A.~Morpurgo,
M.~Craciun, and S.~Margadonna for correspondence. 
This work was partly supported
by MIUR COFIN No. 2003028141-007, MIUR COFIN No. 2004023199-003, by  FIRB RBAU017S8R operated by INFM,
by MIUR FIRB RBAU017S8 R004, and by INFM (Iniziativa trasversale calcolo
parallelo).

%KS-eigenvalues spectra
\begin{table}[p]
\begin{tabular}{ccc}
level & KS-eigenvalue [eV] & symmetry \\ \hline
LUMO+1   & -2.241          & $b_{1u}$ \\
LUMO     & -3.692          & $e_g$    \\
HOMO     & -5.132          & $a_{1u}$ \\
HOMO-1   & -6.092          & $b_{2g}$ \\
HOMO-2   & -6.446          & $b_{2u}$ \\
HOMO-3,-4& -6.526          & $e_u$    
\end{tabular}
\caption{Kohn-Sham eigenvalues spectra and irreducible representation 
of wave--functions near the HOMO--LUMO gap. Note that $e_g$ and $e_u$ 
orbitals are two--fold degenerated.}
\label{tab-electrons}
\end{table}

\begin{table}
\begin{tabular}{cccc}
\hline bond & DFT Ref. \cite{Liao} & Experiment Ref. \cite{Ruan} & this work \\ \hline
$Mg-N_1 $   &   2.008 & 1.990 & 1.991 \\
$N_1-C_1$   &   1.377 & 1.386 & 1.359 \\
$N_2-C_1$   &   1.335 &       & 1.317 \\
$C_1-C_2$   &   1.465 & 1.411 & 1.445 \\
$C_2-C'_2$  &   1.415 & 1.468 & 1.400 \\
$C_2-C_3$   &   1.395 & 1.400 & 1.379 \\ 
$C_3-C_4$   &   1.397 & 1.399 & 1.381 \\
$C_4-C'_4$  &   1.406 & 1.412 & 1.392 \\
$C-H$       &   1.090 & 1.121 & 1.092 \\
$\theta_{C_1-N_1-C'_1} $ & 109.7 & 109.5 & 109.9 \\
$\theta_{N_2-C_1-N_1}  $ & 127.5 & 125.9 & 127.1 \\
$\theta_{N_1-C_1-C_2}  $ & 108.6 & 108.9 & 108.4 
\end{tabular}
\caption{Structure of MgPc. Bond lengths are in \AA, angles in degrees. 
Notations corresponds to labels on Fig. \ref{fig-mol-scheme}}.
\label{tab-struct}
\end{table}

\end{document}